# PTILE: A framework for the Evaluation of Power Transformer Insulation Life in Electric Power System

Nelson O. Ogbogu, Theophilus C. Madueme, and Emmanuel N. Osegi

*Abstract*—In this paper, a framework is developed for power transformer (Generator Step up Unit) insulation life evaluation (PTILE) study on power system Network. Parameters used for studies include real time sample data obtained from power transformer field studies in the South-South Niger Delta region of Nigeria. It is used for performing simulations over varying number of years. Simulation reports shows a polynomial running time complexity and validates the stochastic Hot Spot theory indicating that the transformers in such region should be replaced sooner due to higher hot spots and transformer loading in such regions.

## I. INTRODUCTION

POWER document Transformer Insulation Life Evaluation (PTILE) is programme is a computational attempt to measure the effectiveness of the insulation in a power transformer over a period of time. It seeks to find out how, when and why these transformers fail before expected life time and by so doing, it provides a worthy decision making scheme to avert the occurrence of power failures and shutdowns in a power systems network.

Several attempts have been made in assessing and evaluating the life expectancy of a power distribution transformer. In [1], a methodology has been proposed that supports stochastic and statistical techniques in assessing transformer insulation life expectancy. Such models can give insight into how a PTILE programme may be designed. The effect of thermal loading in indoor/outdoor MV/LV transformers have been evaluated in [5] with far reaching consequences for indoor transformers. Lee et al [7] developed a simplified model based on the IEEE std. C57.91 guide in order to estimate transformer temperatures and elapsed lives based on certain load assumptions.

A comprehensive model for estimating additional accelerated transformer aging in medium/low-voltage distribution transformers due to plug-in electric vehicle charging have been described in [8]. However, very little research has been done in studying the relative effectiveness of potential PTILE algorithms as well as their implementations. Since PTILE is a computationally intensive process, careful choice and engineering of a PTILE algorithm is essential in ensuring timely simulation experiments in the field. For the site engineer, specific tools and structured algorithms are needed to facilitate this study. Our framework should set the direction in this area of analysis.

In this paper, we present an expanding and reusable Algorithmic framework that may be implemented in any one technical programming language. The algorithm encourages the functional approach to tackling the PTILE programming problem and may also be designed to be object oriented – by including a class structure.

This paper is structured in the following order. Section II discusses the analytical details necessary for a PTILE Simulation study. Section III provides the algorithmic design of our proposed framework considering the associated physical/environmental priors. Section IV provides some experimental results using developed algorithm and the case study data. Finally, we give our concluding remarks in Section V.

## II. POWER TRANSFORMER LIFE EVALUATION (PTILE) ANALYTIC FORMULATIONS

The analysis of PTILE will typically include several key factors. Prominent among them are the load (power) consumption rate and the thermal effects in the transformer top-oil, winding as well as changes in the environment. Also, there is fundamental evidence that transformer insulation life may be influenced by the frequency of operation and size of the transformer [4], [6]. In this section, we analyze PTILE based on the hourly load variation and the thermal effects influencing transformer aging.

### A. Load Simulation Study

Load studies will include the determination of a parameter called the load-factor (k) [1], [2], [3] and is expressed as:

N. O. Ogbogu is with the Department of Electrical Engineering,University of Portharcourt, Rivers State, Nigeria (odinaka.ogbogu@uniport.edu.ng).
T. C. Madueme is with the Department of Electrical Engineering,University of Nigeria, Nsukka, Enugu State, Nigeria (theophilus.madueme@unn.edu.ng).
N. E. Osegi is a Graduate student with the Department of Information and Communication Technology, National Open University of Nigeria, Lagos State, Nigeria (nd@osegi.com).

$$k = \frac{I_{actual}}{I_{rated}} \quad (1)$$

Where, $I_{actual}$ is the actual current on load and $I_{rated}$ is transformer name-plate load current. k-factor is an important parameter that may be used to approximate the top-oil temperature rise and the winding hottest temperature rise over top-oil temperature [3].

*B. Step-by-Step Computations for Hourly Load Simulations*

To perform a load simulation study random or varying load data is needed. The load data may be generated analytically by using a pseudo-random number generator, linear step-functions and specialized random algorithms with field minimum and maximum obtained from a specified generator step-up unit (GSU) [1]. These load field values for analytic simulations typically serve as a guide and may include any one of the following:
- The minimum and maximum load currents typically occurring over a specified time range
- The mean load currents over the specified time range
- The mode of the load currents i.e. the most frequently occurring load currents for an integer class of load currents

The specified time range is essentially governed by the Algorithmic Clock (ALGoC) - A pseudo-code approximation of a Natural clock (NC). The ALGoC time base considers at the bare minimum the minutes, hours, days and years of an NC. For the purposes of this study, we shall constrain ourselves to the items i and ii.

Another approach though more expensive is to use the entire field data probably using Graphical-Processing-Units (GPU's) and some deep learning in tandem but this is beyond the scope of this study.

Using the generated or specified load field data, we then pre-compute the load current recursively for the specified number of years as:

$$I_i = \{I_{i1}, I_{i2}, I_{i3}.....I_{i4}\} \quad (2)$$

from the mean and standard deviation of the load currents given in (3) and (4) respectively:

$$\mu_m = \frac{\sum_{i=1}^{n}(I_i)}{n} \quad (3)$$

$$\sigma = \sqrt{\frac{(I_i(i) - \mu_m)^2}{n}} \quad (4)$$

where $I_i$ = a uniform random set of real number load currents
  $\mu_m$ = sample load mean
  $\sigma$ = sample load variance
The sub-optimal values of $I_i$ may then be represented as:

$$I_i^* = \triangleleft I_i \quad (5)$$

for which we expect the system k-factor to be a minimum.

In this way we may perform back-substitution and study the k-values for which $I_{mean}$ is conditionally within range as:

$$k_n(k^* > k_{min}^* \, \& \, k^* < k_{max}^*) = 1 \quad (6)$$

Using (6) we obtain a class of Sub-optimal k-factors which is a function of a normally distributed set of pseudo-random currents.

*C. Hourly Temperature and Life Computation*

To estimate the life of a transformer and its associated parameters we need to compute the hot-spot temperatures for which the top-oil, winding and ambient temperatures seen by the loaded transformer is given.

The net hot-spot temperature given as:

$$\theta_{hs} = \Delta To + \Delta Tw + \Delta Ta \quad (7)$$

is recursively computed over a specified time duration governed by the ALGoC where
  $\Delta To$ = top-oil temperature rise
  $\Delta Tw$ = winding temperature rise and,
  $\Delta Ta$ = ambient temperature rise.

With the exception of the ambient temperature which requires an additional parameter $Z_i$ which is defined as the hourly factors of infinitesimal variations, the temperatures may be computed in a similar manner as in the load case.

*D. Accelerated Aging Factor (FAA) Computation*

The accelerated aging factor typically describes the rate of transformer insulation aging and is defined by the following model:

$$FAA = e^{((\frac{15000}{383}) - (\frac{15000}{\theta_{hs} + 273}))} \quad (8)$$

Life is extended when FAA is less than unity otherwise it is reduced.

*E. Equivalent Aging Factor (FEQA) Computation*

FEQA is a dimensionless quantity that assists in the computation of the ROL. It may be approximated from the relation below:

$$FEQA = \frac{(\sum_{n=1}^{N} FAA_n * \Delta t)}{N * \Delta t} \quad (9)$$

where N = number of years and
  $\Delta t$ = a constant time interval.

The useful life of transformer insulation may be obtained from FEQA subsets or accumulation of FEQA over a given number of years. We consider three interesting points – the mean life, maximum life and minimum life achievable for which a tolerable set of limits may be defined.

ROL is a percentage that describes the potential of transformer failure frequency. Mathematically speaking, ROL may be obtained from the relation:

$$ROL = \frac{life}{k_{RTL}} * 100 \quad (10)$$

where $k_{RTL}$ is a constant representing the normal life for a given transformer based on rate of loss of tensile insulation in transformer.

## III. ALGORITHMIC FRAMEWORK OF PTILE SYSTEM

The proposed Algorithmic framework for determining hourly load variations and transformer thermal characteristics for a PTILE study is given in [9]. The System will include the main-algorithm which requires three sub-algorithms – the Load_function Algorithm, the Temp_function Algorithm and FAA Algorithm with their dependencies. We describe briefly, only the Main, Load_function and Temp_function algorithms here.

### A. Main Algorithm

This requires the load and temperature working parameters to be specified before hand. The core functions Load_Function and Temp_Function are used to obtain the required variables. The Load_Function is used to compute the hourly k-factors while the Temp_function computes FAA, FEQA and the life of transformer from which the ROL may be computed.

### B. Load Algorithm

The load algorithm computes the hourly and hence daily k-factors in accordance with the steps outlined in section II. At the heart of the algorithm is the analytic algorithmic clock (ALGoC), which ensures that the simulations closely resemble real-time observations. The load algorithm requires the rated current, minimum and maximum current as arguments.

### C. Temperature Algorithm

This algorithm has the objective to compute FAA, FEQA and the life of transformer. In order to effectively achieve this objective, analytic clocking ALGoC, is employed. The arguments required are the number of years, minimum and maximum top-oil, winding and ambient temperatures. The Temperature algorithm also has a functional dependency – the FAA algorithm which is easily implemented as defined in (8).

## IV. EXPERIMENTS AND RESULTS

Running time Simulations were performed at intervals of 2years from the 1st running years up to a maximum of 12 years. Life Plots for the 10th, 30th and 50th years are given in Fig 1 to Fig 3 respectively. The rate of loss for 10 and 50 years is also given in Fig 4 and Fig. 5 respectively.

For the running time study, we gradually increase interval time accordingly and record the elapsed time for a complete run for specified number of years. The graphical results show increasing hot-spot and life with increasing year. As can be seen from Figs 1 through 2, the consumption span becomes thinner as the number of years increase showing rapid loss of life. The rate of loss also increases almost linearly with slight perturbations between 40 and 50 years. For 10 year duration a loss of almost 90% is observed (see Fig4).

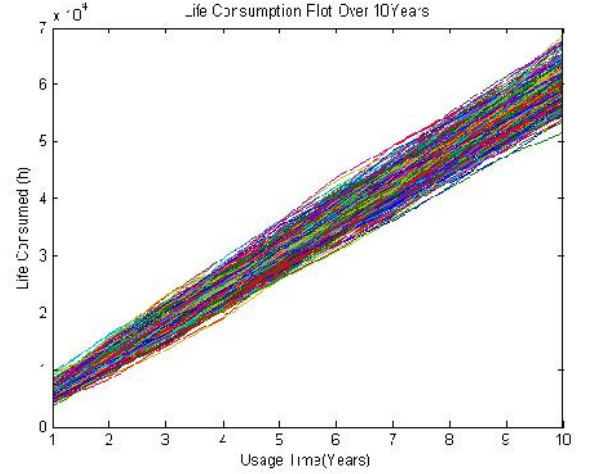

Fig. 1. Life Consumption Plot in hours for 10 years

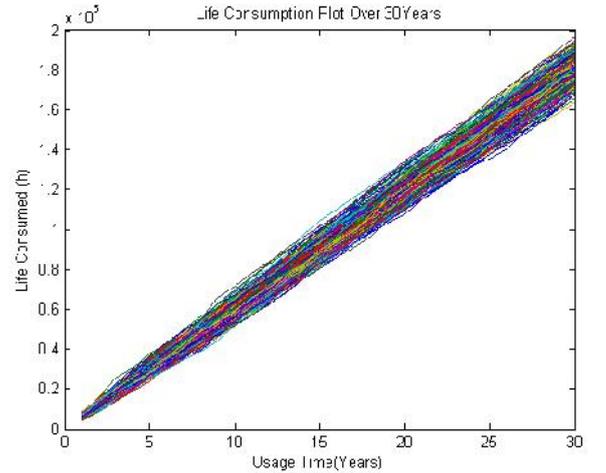

Fig. 2. Life Consumption Plot in hours for 30 years

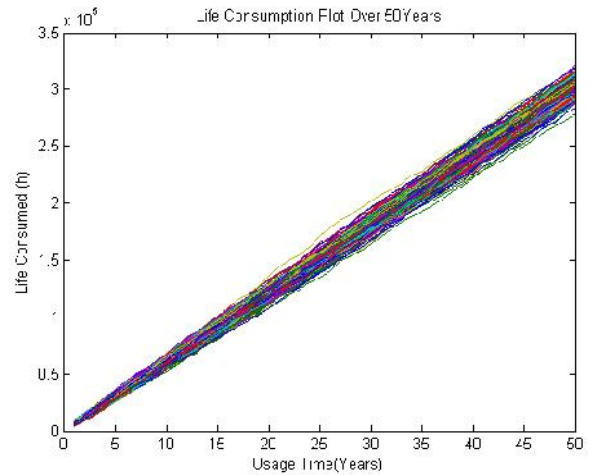

Fig. 3. Life Consumption Plot in hours for 50 years

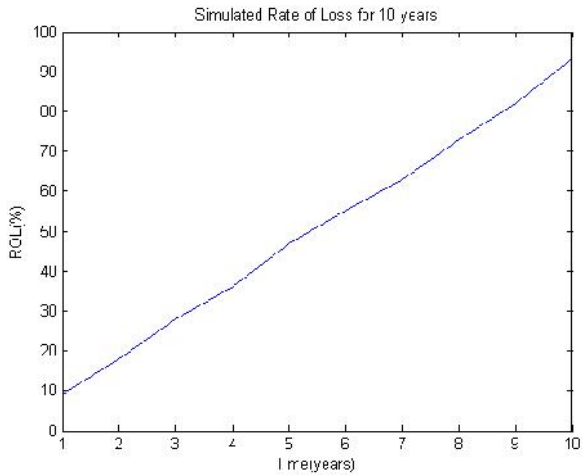

Fig. 4.  Rate of Loss of Transformer Insulation over 10 years

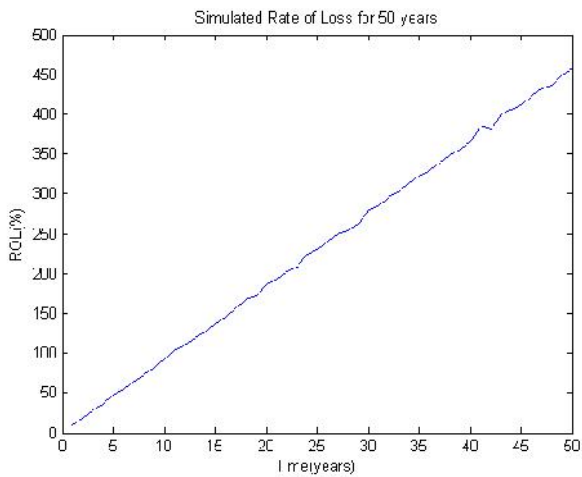

Fig. 5.  Rate of Loss of Transformer Insulation over 10 years

Results of running time simulations are given in Table 1. Running time using analysis of fit at the cubic level (see Fig 6) at estimated root-mean square errors (RMSE) of 1.673 shows an empirical polynomial time complexity performance which is largely attributed to sparsity enforced by 15minutes interval constraint.

TABLE I
RUNNING TIME REPORT FOR 12 YEARS

| S/N | YRS | START TIME (CLOCK P.M.) | END TIME (CLOCK P.M.) | +/- BIAS (S) | TIME INTERVAL (S) |
|---|---|---|---|---|---|
| 1 | 1 | 13:01 | 13.01 | -45s | 45 |
| 2 | 2 | 12:58 | 12:59 |  | 60 |
| 3 | 4 | 13:02 | 13:03 |  | 60 |
| 4 | 6 | 13:05 | 13:07 | +10s | 130 |
| 5 | 8 | 13:09 | 13:12 | +4s | 184 |
| 6 | 10 | 13:14 | 13:18 | -10s | 230 |
| 7 | 12 | 13:19 | 13:24 | -10s | 290 |

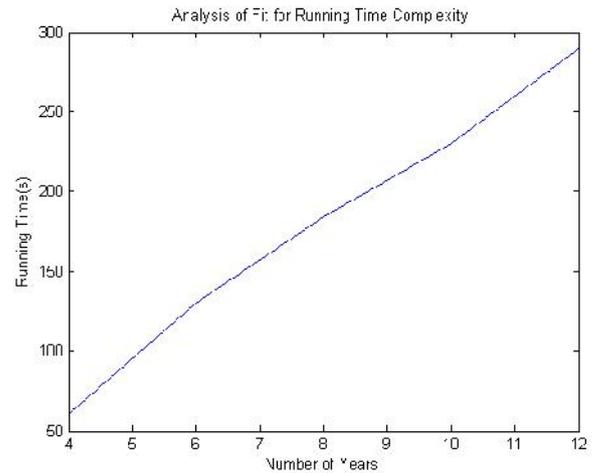

Fig. 6.  Time Complexity Fit for 12 years

## V.  CONCLUSION

The likelihood of transformer failures is a function of several factors of which the temperature and load current combined, plays a vital role in non-temperate regions.  Since transformer insulation life effects are random in nature, better models are needed that fully capture this scenario. In this regard, we developed a framework using a time stamp, to allow insulation life studies to be performed on real live data sets. We have shown using our framework for the case of South-South Nigerian region that insulation life will be decreased at higher hot spot meaning higher ROL's for the given transformer and measures should therefore be taken to replace transformers sooner to avoid total power shutdowns. In addition, mitigation measures by way of load shedding, high quality core windings and special coolants can help increase the life expectancy of transformers.